\documentclass[12pt]{article}
\usepackage{latexsym,graphicx,multirow}
\usepackage{float}
\usepackage{amssymb}
\usepackage{amsmath}
\usepackage{amscd}
\usepackage{amsthm}
\usepackage[left=2cm,top=2.5cm,right=2.5cm,bottom=1.5cm]{geometry}
\usepackage{hyperref}
\usepackage{epstopdf}
\usepackage{cite}

\begin{document}
	
	\begin{center}
		\large{\bf{Statefinder hierarchy model for the Barrow holographic dark energy}} \\
		\vspace{10mm}
		\normalsize{ Vinod Kumar Bhardwaj$^1$, Archana Dixit$^2$ and Anirudh Pradhan$^3$  }\\
		\vspace{5mm}
		\normalsize{$^{1,2,3}$Department of Mathematics, Institute of Applied Sciences and Humanities, GLA University,\\
			Mathura-281 406, Uttar Pradesh, India}\\
		\vspace{2mm}
		$^1$E-mail: dr.vinodbhardwaj@gmail.com\\
		\vspace{2mm}
               $^2$E-mail: archana.dixit@gla.ac.in\\
               \vspace{2mm}
               $^3$E-mail: pradhan.anirudh@gmail.com \\
			
		\vspace{10mm}
	
		 \end{center}
              \begin{abstract}
	
	In this paper, we have used the state finder hierarchy for Barrow holographic dark energy (BHDE) model in the system of the 
	FLRW Universe. We apply two DE diagnostic tools to determine $\Lambda BHDE$ model to have various estimations 
	of $\triangle$. The first diagnostic tool is the state finder hierarchy in which we have studied 
	$S^{(1)}_{3}$  $S^{(1)}_{4}$, $S^{(2)}_{3}$  $S^{(2)}_{4}$ 
	and second  is the  composite null diagnostic (CND) in which the trajectories ( $S^{(1)}_{3} - \epsilon$), ($S^{(1)}_{4}-\epsilon$), 
	($S^{(2)}_{3}-\epsilon$)($S^{(2)}_{4}-\epsilon$) are discussed, where $\epsilon$ is fractional growth parameter. The infrared cut-off 
	here is taken from the Hubble horizon. Finally, together with the growth rate of matter perturbation, we plot the state finder hierarchy that 
	determines a composite null diagnosis that can differentiate emerging dark energy from $\Lambda CDM$. In addition, the combination of 
	the hierarchy of the state finder and the fractional growth parameter has been shown to be a useful tool for diagnosing BHDE, particularly 
	for breaking the degeneracy with different parameter values of the model.
      
      \end{abstract}
 
 
  
\section{Introduction}
The cosmological findings \cite{ref1}-\cite{ref4} have indicated that there is an accelerated expansion. 
The responsible cause behind this accelerated expansion is a miscellaneous element having exotic negative pressure termed as dark energy (DE).
In this field, an enormous DE models are suggested for the conceivable existence of DE such as $\Lambda$CDM models, holographic dark energy (HDE) 
models \cite{ref5}-\cite{ref10}, and the scalar-field models \cite{ref11}-\cite{ref16}, etc. 
In the direction of modified gravity, various efforts have been made, for instance, the $f(R)$  theories \cite{ref16}-\cite{ref17a}, 
the Dvali-Gabadadze-Porrati (DGP) braneworld model \cite{ref18} etc. Amidst other prevailing theoretical perspectives, the least complicated 
amongst all is the $\Lambda$CDM model which is in good agreement with observations. The $\Lambda$CDM model normally suffers from many theoretical 
problems such as fine tuning and coincidence problems \cite{ref19}-\cite{ref22}.\\

For clarifying the accelerating expansion of universe, the dark energy has been considered as most encouraging factor \cite{ref19,ref22,ref23, ref24}. 
A large number of models have been proposed on idea of dark energy, but DE is still assumed as mysterious causes \cite{ref25}-\cite{ref29}. 
Fundamentally, DE problem might be an issue of quantum gravity. It is normally accepted that the holographic principle (HP) \cite{ref30,ref31} 
is one of the fundamental principles in quantum gravity. In 2004, Li proposed the  (HDE) model \cite{ref32}, which is the primary DE model inspired 
by the HP. This model is in generally excellent concurrence with recent observational data \cite{ref33}-\cite{ref38a}. Recently, 
Nojiri et al. \cite{ref38b} proposed a HDE model in which they have established the unification of holographic inflation with holographic dark energy.  \\	

Motivated by the holographic fundamentals and utilizing different framework entropies, some new types of DE models were suggested such as, 
the Tsallis holographic dark energy (THDE) \cite {ref39,ref39a}, the Tsallis agegraphic dark energy (TADE) \cite{ref40} the Renyi holographic dark energy 
(RHDE)  \cite{ref41,ref41a},and the Sharma–Mittal holographic dark energy (SMHDE) model \cite{ref42,ref42a}. Recently many authors shows a great interest in 
HDE models and explored in different context \cite{ref43}-\cite{ref48}. \\
	
Now a day our main concern is the $\Lambda$CDM instead of DE models, 
because our observational analysis are based on $\Lambda$CDM. The $\Lambda HDE$ models generally deal with two key parameters: 
dimensionless HDE parameter and fractional density of cosmological constant. Since they decide the dynamical nature of DE and the 
predetermination of universe. For the evaluation of the various DE models, two diagnostic tools are generally used. The first one is the state finder 
hierarchy \cite{ref49,ref50}, which is a geometrical diagnostic tool and it is model independent. The second parameter is the fractional growth 
parameter (FGP) $\epsilon$ \cite{ref51,ref52}, which presents an independent scale diagnosis of the universe's growth history. Additionally, a variation 
of the state finder hierarchy along with FGP, referred to as composite null diagnosis (CND)\cite{ref50} is often used to diagnose 
DE models.\\

In this way a clearer diagnosis called as state finder hierarchy (SFH) $S_{n}$ has been recently implemented in \cite{ref53}. The $O_{m}$ 
diagnostic and statefinders are connected to the derivative of scale factor $a$ and the expansion rate $H(z)$. Now the composite null 
diagnosis (CND) is a helpful and beneficial technique to the state finder progressive system, where the fractional growth parameter 
$\epsilon$ associated with the structure  growth rate \cite{ref51,ref52}. In this model \cite{ref54}, the researchers developed Barrow 
holographic energy. Here  the authors utilized the holographic principle in a cosmological structure and Barrow entropy rather then the 
popular Bekenstein-Hawking. BHDE is also an engaging and thought-provoking optional framework for quantitatively describing DE.\\
 
Barrow \cite{ref55} has recently found the possibility that the surface of the black hole could have a complex structure down to arbitrarily tiny
due to quantum-gravitational effects. The above potential impacts of the quantum-gravitational space time form on the horizon region
would therefore prompt another black hole entropy relation, the basic concept of black hole thermodynamics. In particular 
 \begin{equation}
 \label{1}
 S_{B}= \left(\frac{B}{B_{0}}\right)^{1+\frac{\triangle}{2}}. 
 \end{equation}
Here $B$ and $B_{0}$ stand for the normal horizon area and the Planck area respectively. The new exponent $\triangle$ is the 
quantum-gravitational deformation with bound as $0 \le \triangle \le 1$ \cite{ref54}-\cite{ref57}. The value $\triangle = 1$ gives 
to the most complex and fractal structure, while $\triangle = 0$ relates to the easiest horizon structure. 
Here as a special case the standard Bekenstein-Hawking entropy is reestablished and the scenario of BHDE has been developed. The  inequality 
$\rho_{_{D}}L^{4}\leqslant S$, is given by the standard HDE, where $L$ stands for the horizon length under the assumption $S \propto A \propto L^{-2}$ 
\cite{ref58}. The Barrow holographic dark energy models have been explored and discussed by various researchers \cite{ref58a} -\cite{ref58d} 
in various other contexts. The above relationship results in some fascinating holographic and cosmological setup results \cite{ref30,ref40}. 
It should be noted that the above relationship offers the usual HDE for the special case of $\triangle = 0$, i.e. $\rho{ {D}} \propto L^{-2}$.
Consequently, the BHDE is definitely a more general paradigm than the standard HDE scenario. We are concentrating here on the general 
($\triangle > 0$) scenario. We consider $H^{-1}$ Hubble horizon (HH) as an IR cut-off ($L$). The energy density of BHDE is expressed as  

\begin{equation}
\label{2}
\rho_{_{D}}= C H^{2-\triangle}
\end{equation}
where $C$ is an unknown parameter.\\

In the current research article, we are considering the fundamental geometry of the universe to be a spatially flat, homogeneous and 
isotropic space time. The Hubble horizon has been regarded as an effective IR cut-off to describe the continuing accelerated expansion 
of the universe \cite{ref54,ref54a}. The objective of this research is to apply the diagnostic tools to differentiate between the BHDE 
models with various other values of $\triangle$. 
Here, we use the diagnostic tool of the state finder hierarchy for BHDE that achieves the value of the $\Lambda CDM$
model and demonstrates consistency of the model for proper estimation of the parameters. It is worth mentioning here that we have 
also demonstrated DE's physical scenario by taking Barrow exponent $\triangle>1$.
We arrange the manuscript as follows: The BHDE model suggested in \cite{ref54} and the basic field equations are added in Section $2$. 
The state finder hierarchy are discussed in Sec. $3$. We explore the fractional state finder diagnostic in Sec. $4$. Growth rate 
perturbations are discussed in Sec. $5$. Section $6$ contains the conclusive statements, discussion and discourse.


\section {Basic field equations} 
The construction of modified Friedman equations with the application of gravity-thermodynamic conjecture is discussed in this section with the 
use of entropy from Barrow \cite{ref59}. The metric reads as for the flat FRW Universe 

\begin{equation}
\label{3}
ds^{2}=-dt^{2}+a^{2}(t)(dr^{2}+r^{2}d\Omega^{2}),
\end{equation}
where $ \Omega^{2}= d\theta^{2} + \sin^{2}\theta d\phi^{2}$, $t$ is the cosmic time and $a(t)$ is the dimensionless scale factor 
normalized to unity at the present time i.e. $a(t_{0}) = 1$. \\

 The Friedmann first field equation for BHDE is given as :
\begin{equation}
\label{4}
H^{2}=\frac{1}{3}8\pi G (\rho_{_{D}}+\rho_{m}),
\end{equation}
where $ \rho_{_{D}}$ and $\rho_{m}$ read as the energy density of BHDE and matter and expressed as  $\Omega_{_{D}}= \frac{8\pi\rho_{_{D}}G}{3 H^{2}}$ 
and $\Omega_{m}= \frac{8\pi\rho_{m}G}{3 H^{2}}$ respectively. The conservation laws for matter and BHDE are defined as:
$\dot\rho_{m}+3H\rho_{m}=0, \dot\rho_{_{D}}+3H(p_{_{D}}+\rho_{_{D}})=0$.\\
 
From Eq (\ref{2}), we get
 \begin{equation}
 \label{5}
 \dot\rho_{_{D}}= \frac{3C}{2} (2-\triangle)H^{2-\triangle} \left(\frac{\Delta \Omega _D}{(\Delta -2) \Omega _D+2}-1\right).
 \end{equation}
 Now, using  Eq (\ref{4}), we obtain
 \begin{equation}
 \label{6}
 \frac{\dot{H}}{H^2}=\frac{3}{2} \left(\frac{\Delta \Omega _D}{(\Delta -2) \Omega _D+2}-1\right).
 \end{equation}
Using  Eq. (\ref{6}), the deceleration parameter (DP) $q$ is expressed by 
\begin{equation}
 \label{7}
q= -1-\frac{\dot H}{H^{2}} = \frac{1-(\Delta +1) \Omega _D}{(\Delta -2) \Omega _D+2}.
 \end{equation}
By utilizing the Eqs. (\ref{5}) and (\ref{6}), we get  the EoS parameter  as:
\begin{equation}
\label{8}
\omega_{D}=-\frac{\Delta }{(\Delta -2) \Omega _D+2}.
\end{equation}
From Eqs. (\ref{6}) and (\ref{8}), we derive $\omega_{_{D}}^{'}$ as
\begin{equation}
 \label{9}
  \omega^{'}_D= -\frac{3 \Delta ^3 \left(\Omega _D-1\right) \Omega _D}{\left((\Delta -2) \Omega _D+2\right){}^3},
 \end{equation}
where dash is the derivative with respect to $lna$.
In similar way by the use of Eqs. (\ref{6}) and (\ref{8}), we obtain $\Omega^{'}_{D}$ as:

 \begin{equation}
 \label{10}
 \Omega^{'}_D=-\frac{3 \Delta  \Omega _D \left(\Omega _D-1\right)}{(\Delta -2) \Omega_{D} +2}.
 \end{equation}


\section{Statefinder hierarchy}
Statefinder hierarchy \cite{ref60,ref61,ref62} is an effective diagnosis of geometry. It is appropriated to discriminate, various DE models 
from the $\Lambda$CDM model by utilizes the higher order derivative of the scale factor. In this case, the scale factor a(t) is the main dynamical 
variable. Here we will be focused on the late time evolution of the universe. Now we assume the Taylor expansion of the scale factor around 
the current age $t_{0}$\cite{ref53}:
\begin{equation}
\label{11}
(1+z)^{-1}=\frac{a(t)}{a_{0}}=1+\sum^{\infty}_{n=1} \frac{A_{n}(t_{0})}{n!} \left[H_{0} (t-t_{0})\right]^2
\end{equation}
where  $A_{n}= \frac{a^{n}}{a H^{n}}$~~$n\in N$.
$a^{n}$ is the $n^{th}$ derivative of the scale factor with respect
to time. The  deceleration parameter  $q\equiv- A_{2}$,
$A_{3}$, has been known as the Statefinder ($r$) \cite{ref63} similar as the jerk (j) \cite{ref64}. $A_{4}$ is the snap (s) and $A_{5}$ 
is the lerk (l) \cite{ref65,ref66,ref67}. It is very surprising that, in a spatially flat universe, comprising of pressure less issue and a 
cosmological constant relate to the  $\Lambda CDM$, while we shall represents all the parameters of $ A_{n}$ as the function of $q$ and 
density parameter $\Omega_{m}$. Likewise\\

$ A_{2}=1-\frac{3}{2} \Omega_m $, \\  $ A_3 =1 $, \\
$ A_4 = 1-\frac{3^2}{2} \Omega_m $, \\ 
$ A_5 = 1+3 \Omega_m+\frac{3^3}{2} \Omega^{2}_m $,\\
$ A_6 = 1-\frac{3^3}{2} \Omega_m-3^4 \Omega^{2}_m-\frac{3^4}{4} \Omega^{3}_m$, etc.,\\
where $\Omega_{m} = \frac{\Omega_{_0m} (1+z)^{3}}{h^{2}(z)}$ and $\Omega_{m} =\frac{2}{3}(1+q)$ is the concordance cosmology. 
It is interesting to note that the above expressions leads to define the Statefinder hierarchy $S_{n}$:\\
$ S_2=A_2+\frac{3}{2} \Omega_m $,\\  $S_3=A_3 $, \\
$ S_4 = A_4+\frac{3^2}{2} \Omega_m $, \\ 
$ S_5 = A_5-3 \Omega_m-\frac{3^3}{2} \Omega^{2}_m $,\\
$ S_6 = A_6 +\frac{3^3}{2} \Omega_m+3^4 \Omega^{2}_m+\frac{3^4}{4} \Omega^{3}_m$, etc.

The fundamental characteristic of this diagnostic, that all the $S_{n}$ parameters remains fixed at unity for $\Lambda CDM $ model 
while the astronomical development is going on. $S_{n}\mid_{\Lambda CDM} = 1$ describe  the null diagnostic for concordance cosmology.

\section{ Fractional Statefinder}

The statefinder hierarchy equations produce a progression of diagnostic for $\Lambda CDM$ model with $n>3$. By using the relationship 
$\Omega_{m} = \frac{2}{3}(1+q)$, valid in $\Lambda$CDM. It is also possible to write Statefinders in an alternative form as:\\ 

$ S^{(1)}_3 = A_3 $,\\
$ S^{(1)}_4 = A_4+3 (q+1) $,\\
$ S^{(1)}_5 = A_5-2 (4+3q) (1+q) $, etc \\
It is clearly visible that in both of the methods yield identical results for $\Lambda CDM$: $ S^{(1)}_4= S_{4}$, $ S^{(1)}_5= S_{5}$ =1. 
However for other DE models, it is proposed that the two alternative definitions, of the  state finder, $S_{n}\&S^{1}_{n}$, would provide 
different results. It was shown in \cite{ref49} that a second State finder could be used as $ S^{(1)}_3 = S_3 $\\
 \begin{equation}
 \label{12}
 S^{(2)}_3 = \frac{S^{(1)}_3 -1}{3 (q-\frac{1}{2})}
\end{equation}
In concordance cosmology $S^{(1)}_{3}=1$, while $S^{(2)}_{3}$ fix at zero. As a consequently the Statefinder pair $(S^{(1)}_{3}, S^{(2)}_{3})={(1,0)}$ 
for $\Lambda CDM$ model. Similarly we characterize the second member of the Statefinder hierarchy as
follows
\begin{equation}
\label{13}
 S^{(2)}_n = \frac{S^{(1)}_n -1}{\zeta (q-\frac{1}{2})}.
\end{equation}
Here $\zeta$ is the arbitrary constant.\\
 
In concordance cosmology
$S^{(2)}_{n} =0$ for $\Lambda CDM $.
$\left\{S^{(1)}_n, S^{(2)}_n\right\}=\left\{1,0\right\}$.
The second statefinder $S^{(2)}_{n}$ helps in the desired aim of creating a break in some of the degeneracies present in $S^{(1)}_{n}$. 
We give the specific expression of $S^{(1)}_{3}$, $S^{(2)}_{3}$,$S^{(1)}_{4}$, $S^{(2)}_{4}$ by using the variable $\Omega_{D}$ and 
$\omega_{D}$ depending on the redshift as

\begin{equation}
\label{14}
S^{(1)}_3=1+\frac{9}{2} \omega_{D} \Omega_{D} (1+\omega_{D}) =1-\frac{9 (\Delta -2) \Delta  \left(\Omega_{D}-1\right) \Omega _D}{2 \left((\Delta -2) 
\Omega_{D}+2\right){}^2},
\end{equation} 
\begin{equation}
\label{15}
S^{(2)}_3=1+\omega_{D}=1-\frac{\Delta }{(\Delta -2) \Omega_{D}+2},
\end{equation}
\begin{eqnarray}
\label{16}
& S^{(1)}_4=-\frac{1}{4} 27 \left(\omega _D+1\right) \left(\omega _D \Omega _D\right){}^2+\frac{27}{2} \omega _D \left(\omega _D+1\right) 
\left(\omega _D+\frac{7}{6}\right) \Omega _D+1, \nonumber\\
&=\frac{-\left(43 \Delta ^3-141 \Delta ^2+102 \Delta +16\right) \Omega _D^3+6 \left(12 \Delta ^3-43 \Delta ^2+34 \Delta +8\right) \Omega _D^2-3 
\left(9 \Delta ^3-39 \Delta ^2+34 \Delta +16\right) \Omega _D+16}{2 \left((\Delta -2) \Omega _D+2\right){}^3},
\end{eqnarray} 
\begin{eqnarray}
\label{17}
& S^{(2)}_4=-\frac{1}{2} \omega _D \left(\omega _D+1\right) \Omega _D-\left(\omega _D+1\right) \left(\omega _D+\frac{7}{6}\right),\nonumber\\
&=-\frac{(\Delta -2) \left(\Omega _D-1\right) \left((2 \Delta -7) \Omega _D-3 \Delta +7\right)}{3 \left((\Delta -2) \Omega _D+2\right){}^2}.
\end{eqnarray}


\begin{figure}[H]
	\centering
	(a)	\includegraphics[width=7.5cm,height=7.5cm,angle=0]{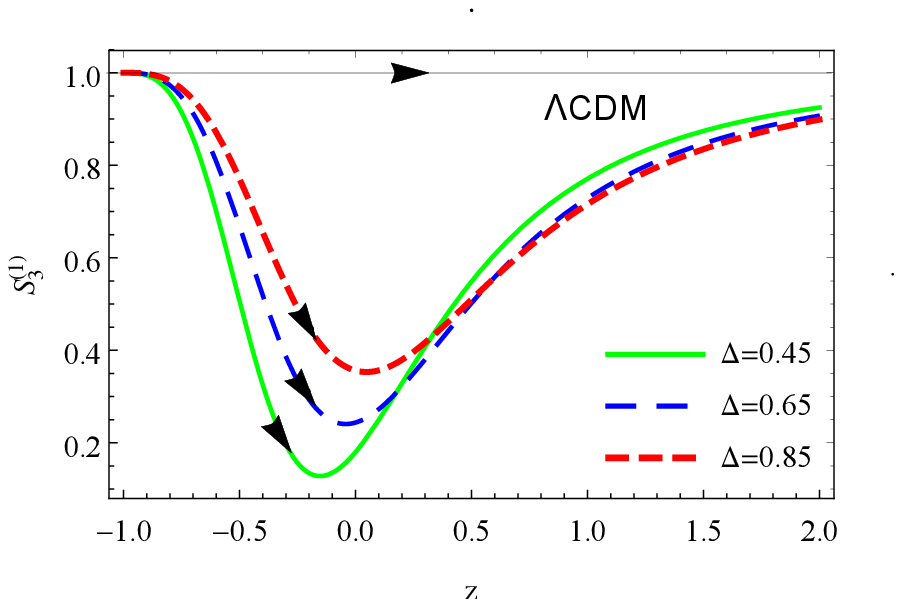}
	(b)\includegraphics[width=7.5cm,height=7.5cm,angle=0]{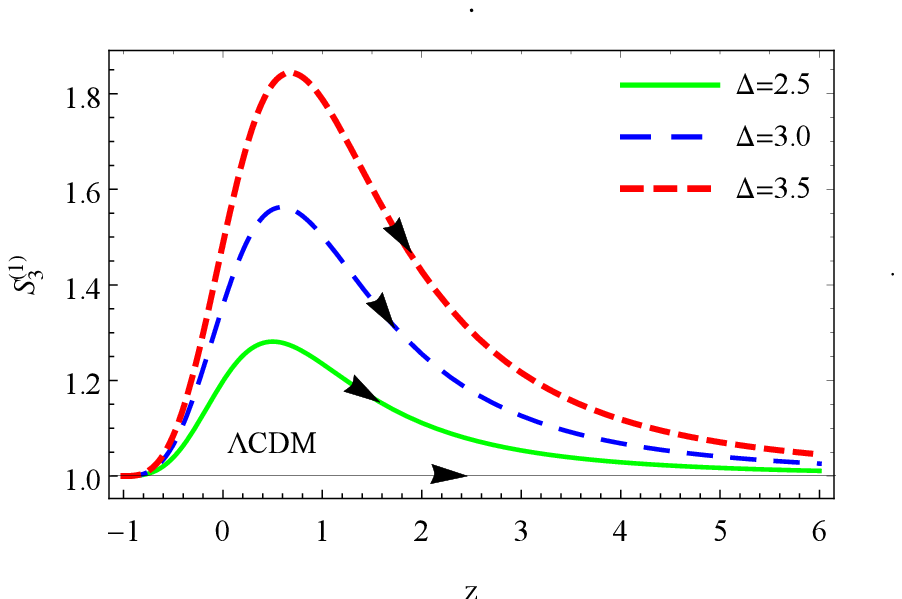}
	\caption{ Evolution trajectories $S_{3}^{(1)}$ verses $z$ for different values of $\triangle$}
\end{figure}

In Figures $1a$ $\&$ $1b$, we plot the evolutionary trajectories of $S^{(1)}_{3}$ verses redshift $z$ for $\Lambda BHDE$ model by using 
$\Omega_{m_{0}}=0.27$, $C=3$ and for different values of Barrow exponent $\triangle$. The curves associated with Barrow exponent $\triangle$ 
and found a shape convex vertex below the $\Lambda CDM$ line for $\triangle<2$ and also found a concave vertex shape in above 
the $\Lambda CDM$ line by  considering $\triangle>2$. The curve has coincidence nature in fig 1a and 1b and again approaches towards 
the $\Lambda CDM$. From the figures $1a$ $\&$ $1b$, we notice that $S_{3}^{(1)}$ evolves decreasingly from $1$ when $\triangle<2$ and $S_{3}^{(1)}$ 
evolves increasing from $1$ if $\triangle>2$. All the curves of $S_{3}^{(1)}$(z) start below the $\Lambda$CDM line  $S_{3}^{(1)}=1$ 
if $\triangle<2$ and crossing the $\Lambda CDM$, line for $\triangle>2$. In order to draw a relevant comparison, the result of 
the $\Lambda CDM$ model is represented additionally by a solid line. The curves of $S_{3}^{(1)}=1$ segregate well from $\Lambda$CDM.

\begin{figure}[H]
	\centering
	(a)	\includegraphics[width=7.5cm,height=7.5cm,angle=0]{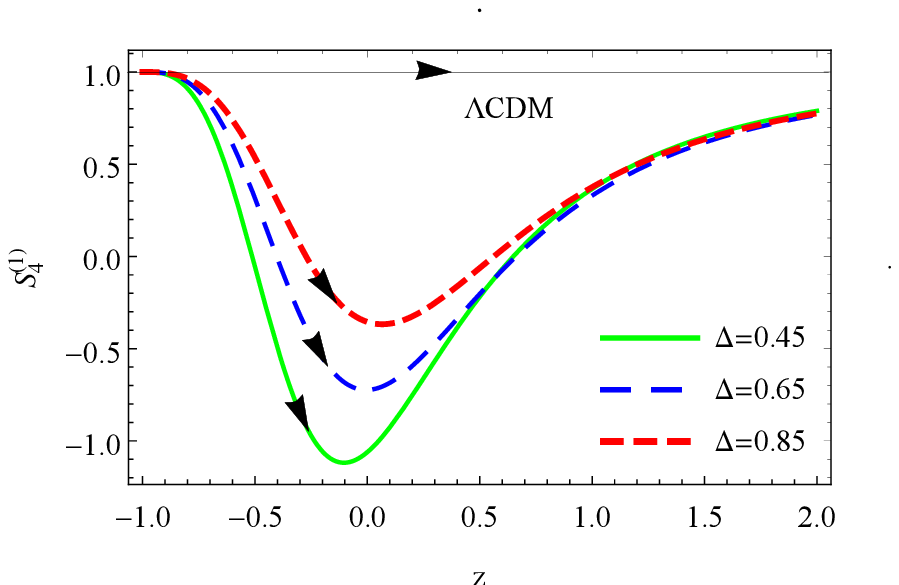}
	(b)\includegraphics[width=7.5cm,height=7.5cm,angle=0]{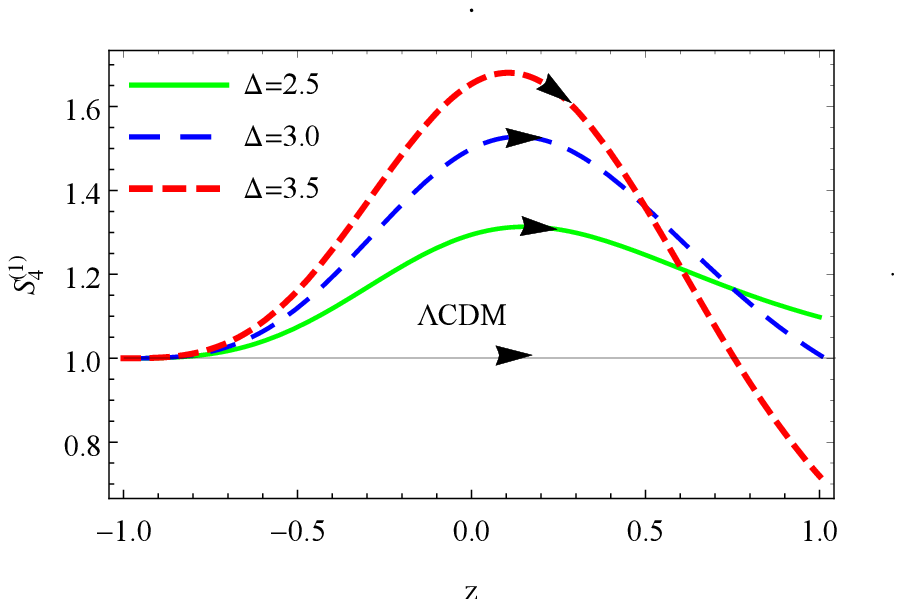}
	\caption{ Evolution trajectories $S_{4}^{(1)}$ verses $z$ for different values of $\triangle$}
\end{figure}
Figures $2a$ $\&$ $2b$ represent the evolutions of $ S^{(1)}_{4}$  versus redshift $z$ for the different values of Barrow exponent 
$\triangle$. The figures depict the behaviour of BHDE and their results are compared with the $\Lambda$CDM  line. 
The evolutionary trajectories of $S^{(1)}_{4}$ show similar behaviour as of $S^{(1)}_{3}$. In Fig. $2a$ trajectories develop below the $\Lambda$CDM 
line $(S_{4}^{(1)}=1)$ near the high redshift region and monotonically decreases then increases and finally tends to the $\Lambda$CDM for 
$\triangle<2 $. Fig. $2b$ depicts the trajectories of $S^{(1)}_{4}$ for $\triangle> 2$ above the $\Lambda$CDM line and form the concave 
vertex at high redshift. These trajectories are significantly separated and crossing the $\Lambda$CDM line.

\begin{figure}[H]
	\centering
	(a)	\includegraphics[width=7.5cm,height=7.5cm,angle=0]{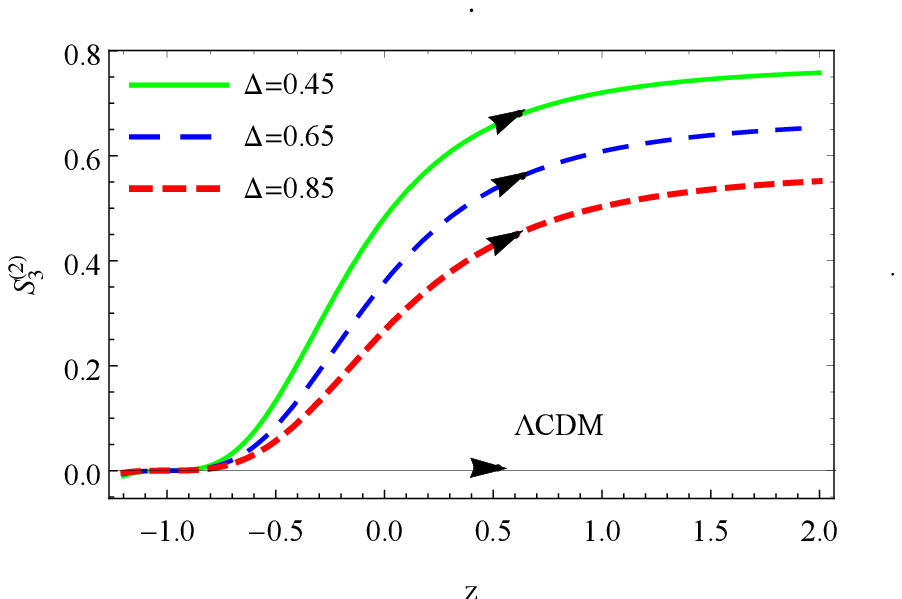}
	(b)\includegraphics[width=7.5cm,height=7.5cm,angle=0]{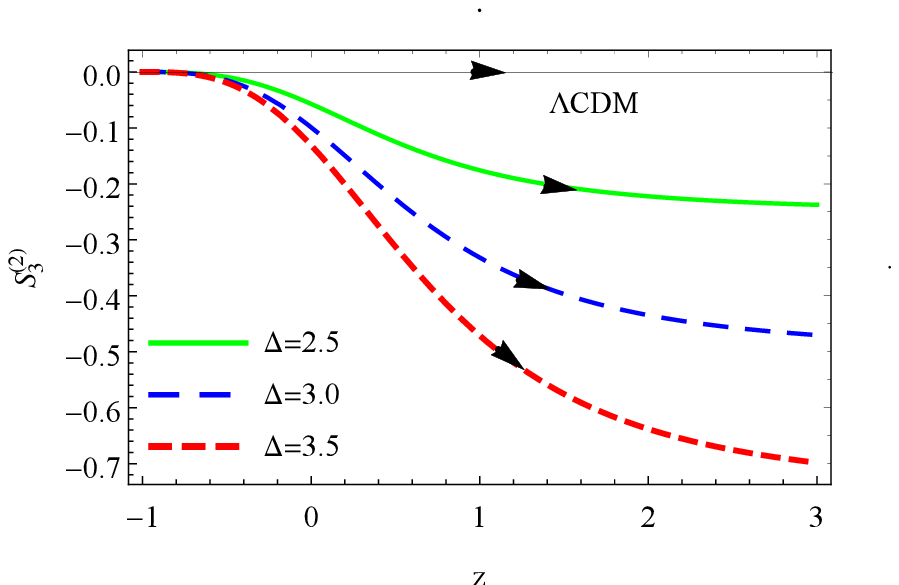}
	\caption{ Evolution trajectories $S_{3}^{(2)}$ verses $z$ for different values of $\triangle$}
\end{figure}

Figure $3a$ shows the evolutionary trajectories of $S^{(2)}_{3}$
for the BHDE model for different values of parameter $\Lambda<2$. 
 We observe that the evolutionary trajectories of $S^{(2)}_{3}$ are distinct in themselves and can be differentiated from $\Lambda CDM$ 
 line  $S^{(2)}_{3}=0$, at high red-shift region for $\triangle<2$.
It evolves the $\Lambda CDM$ line and separated well at high redshift.
Similarly in Fig. $3b$ the trajectories evolves below the $\Lambda$CDM line and separated well in high redshift region, while at low redshift 
these trajectories degenerate closer together.

\begin{figure}[H]
	\centering
	(a)	\includegraphics[width=7.5cm,height=7.5cm,angle=0]{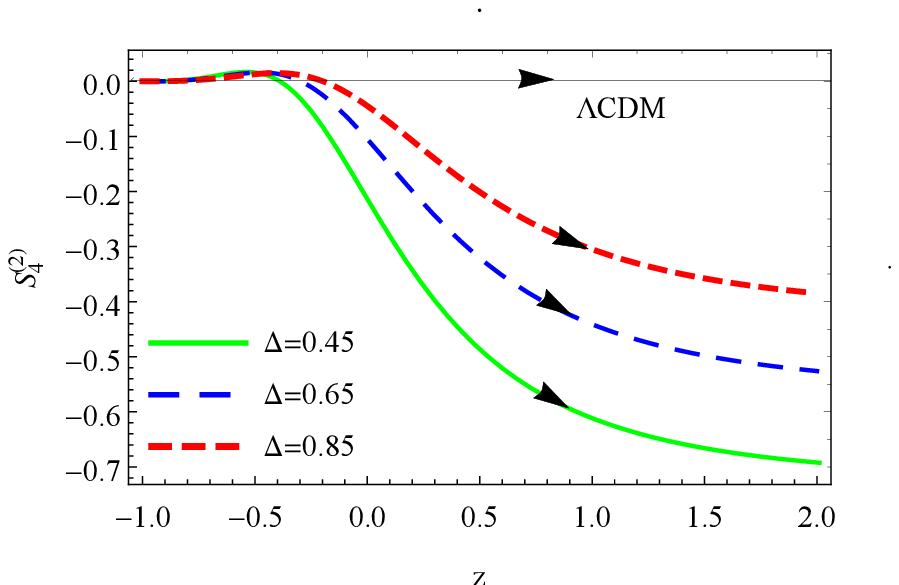}
	(b)\includegraphics[width=7.5cm,height=7.5cm,angle=0]{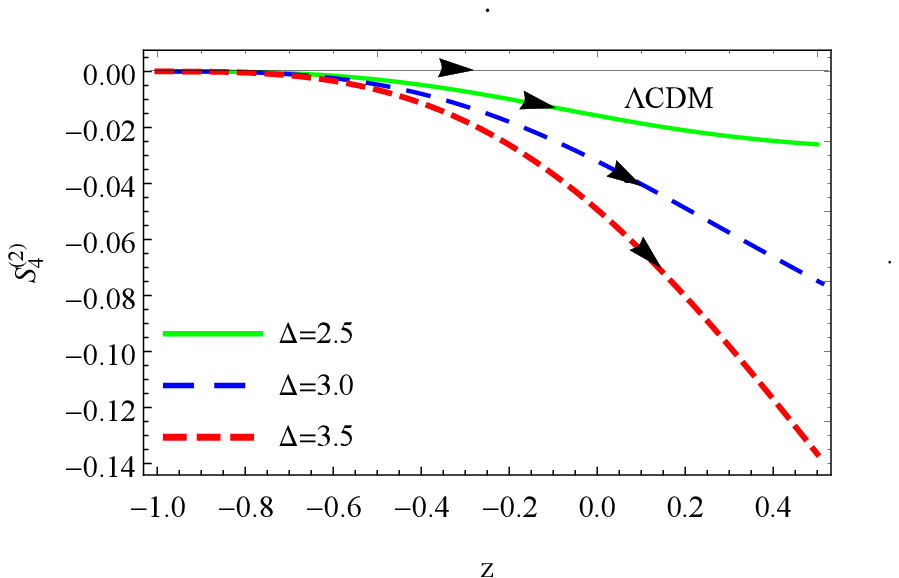}
		\caption{ Evolution trajectories $S_{4}^{(2)}$ verses $z$ for different values of $\triangle$}
\end{figure}

In Figures $4a$ $\&$ $4b$ the graphs show $S^{(2)}_{4}$ evolution versus $z$. These figures are very much separated and evolve
below the $ \Lambda CDM$ line $S^{(2)}_{4}=0$ at the high redshift. These trans-formative directions shows just quantitative effect 
on the $S^{(2)}_{4}$ by fluctuating $\triangle$ and at low redshift the trajectories degenerate closely together with $\Lambda CDM$.

 \begin{figure}[H]
	\centering
	(a)	\includegraphics[width=7.0cm,height=7.0cm,angle=0]{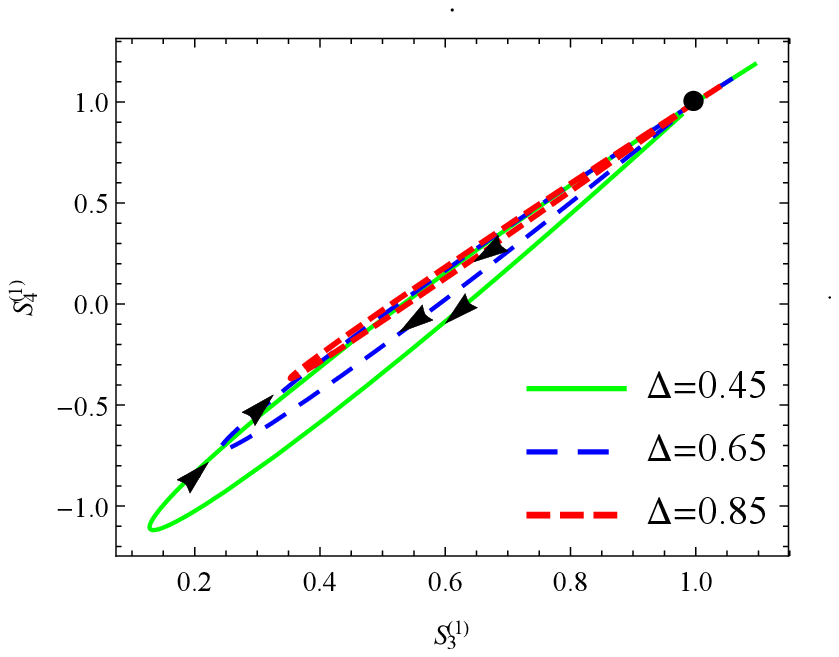}
	(b)\includegraphics[width=7.0cm,height=7.0cm,angle=0]{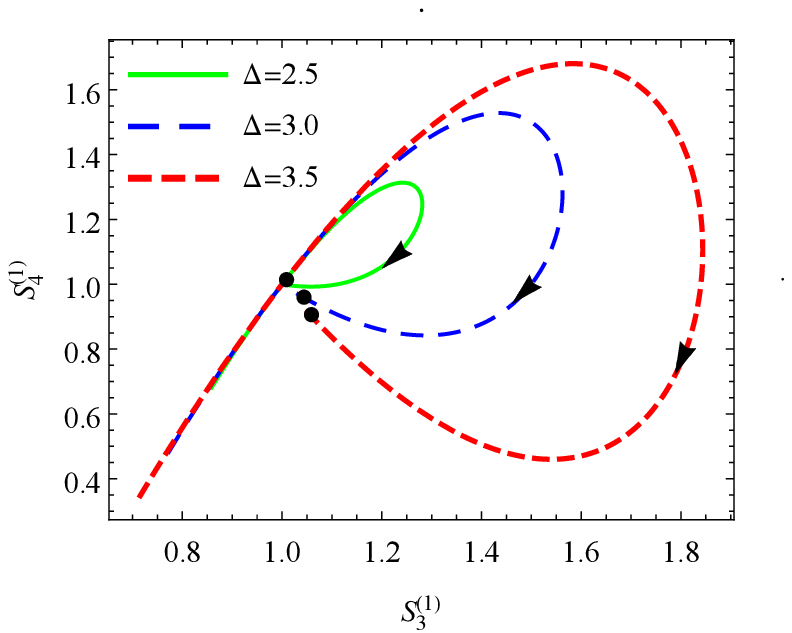}
	\caption{ Evolution trajectories $S_{4}^{(1)}$ verses $S_{3}^{(1)}$  for different values of $\triangle$}
	\end{figure}
	
Figures $5a$ $\&$ $5b$ depict the Statefinders $S_{4}^{(1)}$ and  $S_{3}^{(1)}\equiv S_{3}$
are expressed for BHDE model; the arrows and dots represent time evolution and present epoch respectively by using $\Omega_{m0}=0.27$. 
$\Lambda$CDM relates to fixed point (1,1) for $\triangle>2$ and $\triangle<2$. 


\begin{figure}[H]
	\centering
	(a)	\includegraphics[width=7.5cm,height=7.5cm,angle=0]{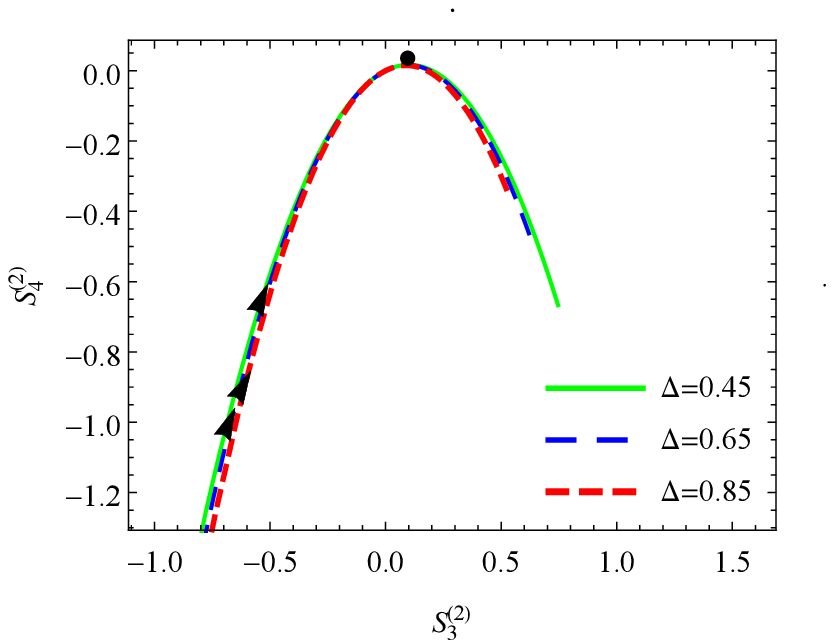}
	(b)\includegraphics[width=7.5cm,height=7.5cm,angle=0]{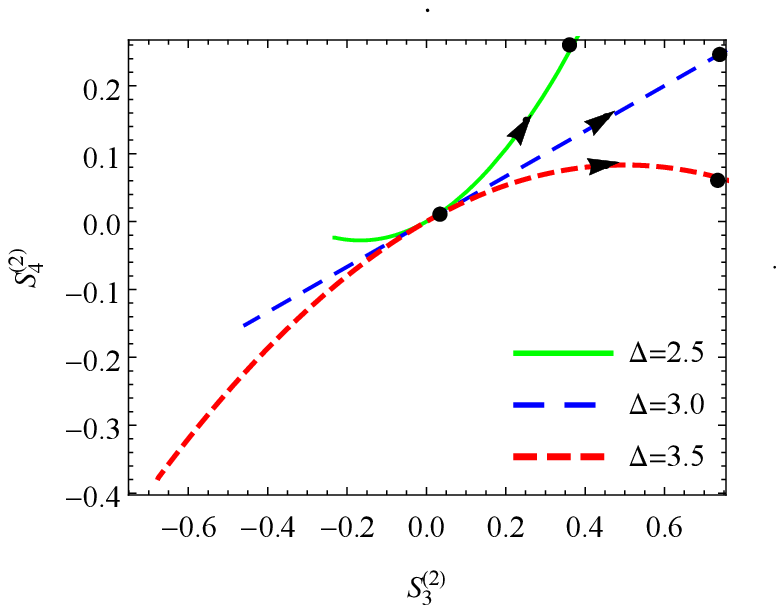}
		\caption{ Evolution trajectories $S_{4}^{(2)}$ verses $S_{3}^{(2)}$  for different values of $\triangle$}
	
\end{figure}

The Statefinders  $S_{4}^{(2)}$  and $S_{3}^{(2)}$ are also shown in Figs. $6a$ $\&$ $6b$ for BHDE model. The fixed point at {(0,0)} in figure 
is $\Lambda CDM$. The arrows show time evolution furthermore, the current age in the various models is appeared as
a dot by using $\Omega_{m0}=0.27$ for $\triangle>2$ and $\triangle<2$. 
As we demonstrate in Figs. $(1-6)$, the Statefinder hierarchy
give us an incredible methods for recognizing dynamical DE models from $\Lambda CDM$.


\section{Growth rate of perturbations}
We use the "parametrized post-Friedmann" theoretical framework for the estimation of growth rate perturbations for
the interaction of DE to acquire the $\epsilon(z)$ values for the models numerically. \\

The perturbation growth rate parameter $\epsilon$ \cite{ref51,ref52} is given by

\begin{equation}
\label{18}
\epsilon =\frac{f(z)}{f_{\Lambda CDM}(z)},
\end{equation}
where $f(z)= \frac{d log\delta}{d loga}$ represents the structure's growth rate. Here, $\delta=\frac{\delta_{\rho_{m}}}{\rho_{m}}$, with 
$\delta_{\rho_{m}} \to$ matter density perturbation,  $\rho_{m} \to$ energy density.\\
 
If the matter density perturbation is linear and there is no correlation
between CDM and DE, the equation of late-time perturbation may be given as
 
 \begin{equation}
 \label{19}
 \ddot \delta + 2H\dot{\delta}= 4\pi G {\rho_{m}}\delta,
 \end{equation}
 where $G \to$ gravitational constant. According for the  linear density perturbation the growth rate is approximatively 
 determined by\cite{ref62}:
 
 \begin{equation}
 \label{20}
 	f(z) =\Omega_{m}(z)^{\gamma},
 \end{equation}

 \begin{equation}
 \label{21}
\gamma =\frac{3}{5-\frac{\omega}{1-\omega}}+\frac{3 (1+\omega) (1-\frac{3}{2}\omega)}{125(1-\frac{6}{5}\omega)^{3}} (1-\Omega_m{(z)}),
\end{equation}
where $\Omega_{m}(z)=\frac{\rho_{m}}{3H^{2}M_{p}^{2}}$ denotes the fractional density of matter. 
For example $\gamma = 0.55$, the above approximation follows the physical DE models well, when $\omega$ is either a constant, or varies
slowly  for $\Lambda CDM$ \cite{ref62,ref68}.\\

In case of models other than this, the values taken up by $\epsilon(z)$ display variations from $\Lambda$CDM for which the parameter 
keeping track of fractional growth is employed as a diagnostic. In the case of an interacting DE model, however, the growth rate can not be 
parametrized simply by Eqs. (\ref{20}) and (\ref{21}) \cite{ref69}. 

This equation can be numerically solved for the condition $f (z = 0) = 1$ for the $\Lambda$CDM  
and the BHDE model. Depending on $f (z)$, other null diagnostic, known as fractional growth parameter $\epsilon$, is discussed in \cite{ref50,ref61}.

\subsection{Analysis of the composite null diagnostic}

In this section, we have  also introduced a composite null diagnostic (CND), which is a combination of members of state finder hierarchy 
and fractional growth parameter. This investigation  of CND can be helpful to represent the conduct of geometrical and matter perturbation 
data of enormous development. In this paper, we use four CND pairs, $(S_{3}^{(1)}, \epsilon)$ and  $(S_{4}^{(1)}, \epsilon)$, 
$(S_{3}^{(2)}, \epsilon)$, and $(S_{4}^{(2)}, \epsilon)$ respectively. Initially we start  the study of the fractional growth parameter 
$\epsilon(z)$ to diagnose the $\Lambda BHDE $ model. The evolutionary trajectories of $(S_{3}^{(1)}, \epsilon)$ for BHDE model are plotted 
in figures $7a$ and $7b$, where the current value of $(S_{3}^{(1)}, \epsilon)$ is labeled with circular dots, and the fixed point 
$S_{3}^{(1)}$ = {(1, 1)} for the $\Lambda CDM$ model. It is also additionally appeared as a star for  comparison. Figures $7a$ and $7b$ plot with 
$S_{3}^{(1)}$ verses $\epsilon$ for different values of $\triangle$. So we can use a composite null diagnostic (CND), as a 
combination of $(S_{3}^{(1)}, \epsilon)$. From the figure we noticed that, curve evolves near $\Lambda CDM$  and monotonically decreases 
and forming convex vertices for $\triangle<2$ and for $\triangle>2$ it forms convex vertices and  approaches to $\Lambda CDM$ at high red shift.
Figs. $7a$ and $7b$ show the evolutionary trajectories of $(S_{3}^{(1)}, \epsilon)$ of BHDE model by considering various estimations of 
$\triangle$, where the star signifies the $\Lambda CDM$ model of $(S_{3}^{(1)}, \epsilon)$. 


\begin{figure}[H]
	\centering
	(a)	\includegraphics[width=7.5cm,height=7.5cm,angle=0]{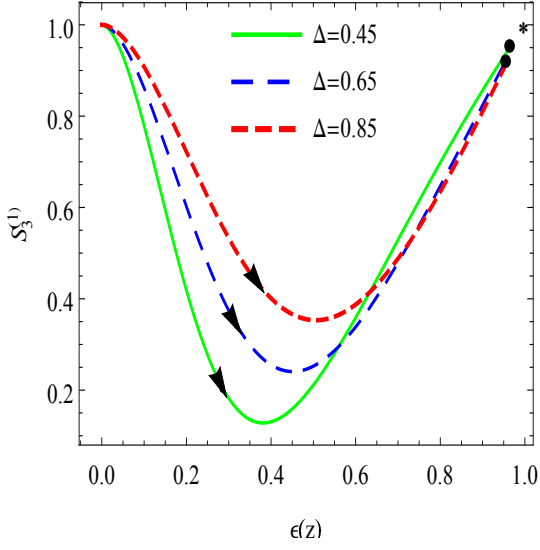}
	(b)\includegraphics[width=7.5cm,height=7.5cm,angle=0]{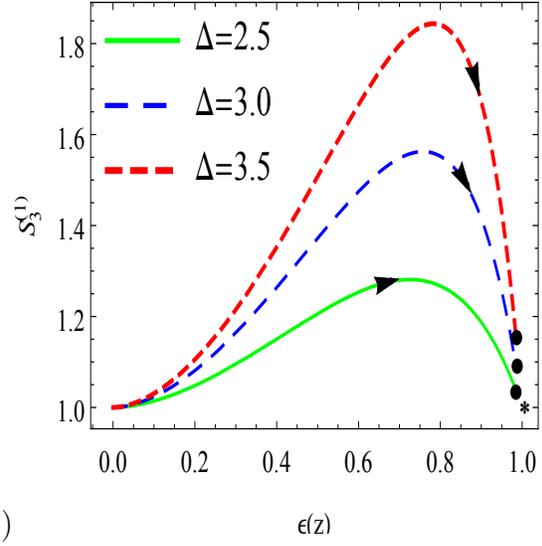}
	\caption{ Evolution trajectories $S_{3}^{(1)}$ verses $\epsilon$ for different values of $\triangle$}
	
\end{figure}
\begin{figure}[H]
	\centering
	(a)	\includegraphics[width=7.5cm,height=7.5cm,angle=0]{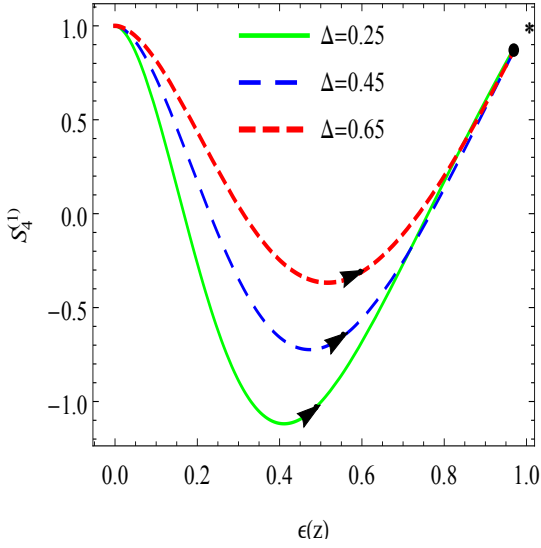}
	(b)\includegraphics[width=7.5cm,height=7.5cm,angle=0]{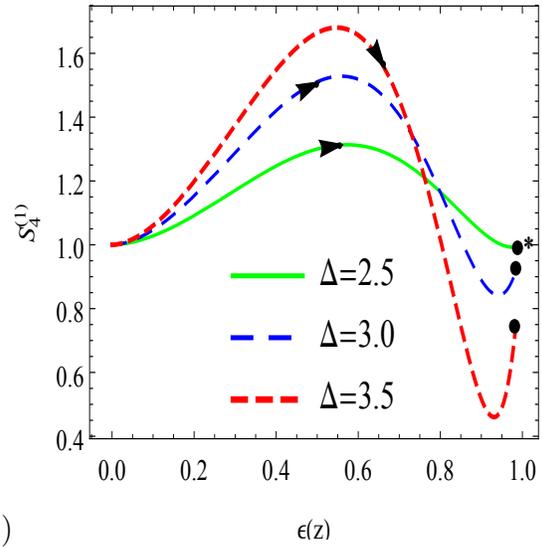}
	\caption{ Evolution trajectories $S_{4}^{(1)}$ verses $\epsilon$ for different values of $\triangle$}
	
\end{figure}

In addition, the CND $(S_{4}^{(1)}-\epsilon)$ is also used to study the BHDE model. We draw the evolutionary pathways for the Barrow holographic 
DE model. Figs. $8a$ $\&$ $8b$ show the present-day values  of $(S_{4}^{(1)}, \epsilon)$  and the fixed point {(1, 1)} for $\Lambda CDM$ and
direct measuring of effective distances between them. The evolutionary trajectories  $(S_{4}^{(1)}-\epsilon)$  forms the convex vertex for 
$\triangle <2$. The current values of $(S_{4}^{(1)},\epsilon)$ for the BHDE are shown by dots. Similarly if we take  $\triangle >2$ initially 
the curves form a concave vertex. The curves of CND pairs $(S_{4}^{(1)}, \epsilon)$ show similar behavior of the curve $(S_{3}^{(1)}, \epsilon)$.

  \section{Conclusion}
  In this manuscript we have discussed the BHDE by utilizing the Barrow entropy, rather than the standard Bekenstein-Hawking. Our  main focus 
  in the manuscript is to diagnose BHDE model by the State finder hierarchy  and composite null diagnostic (CND).
  The  diagnostic tools which are frequently apply to test the DE models: (a) state finder hierarchy  $(S_{3}^{(1)}, \epsilon)$ (b) fractional 
  growth parameter $\epsilon$. Additionally the CND pair, are also considered to diagnosing DE models. So, the major concern of the
 work is to apply diagnostic tools for the BHDE models. \\
 
 As per the following, the principal highlights of the models are :\\

 \begin{itemize}
 \item
  In Figures (1-6), the statefinder hierarchy gives the analytical expressions of $S_{3}^{(1)}$, $S_{4}^{(1)}$, $S_{3}^{(2)}$, $S_{4}^{(2)}$ and 
  elaborate numerically, for BHDE model. We have plotted the evolutionary curves of $S_{3}^{(1)}$, $S_{4}^{(1)}$,
  $S_{3}^{(2)}$, $S_{4}^{(2)}$  with respect to the redshift $z$ and $\epsilon(z)$. The evolutionary trajectories of $S_{3}^{(1)}$, $S_{4}^{(1)}$ evolve
  below the $\Lambda$CDM line and form the convex vertex for the barrow exponent $\triangle<2$  while  $S_{3}^{(1)}$ $S_{4}^{(1)}$ expand from above 
  the $\Lambda$CDM line  and form the concave vertex for $\triangle>2$. 
 
 \item  
 We have also examined the the trajectories  $(S_{3}^{(1)}, \epsilon)$ and $(S_{4}^{(1)}, \epsilon)$ for BHDE. Evolutionary trends
 have similar approach as shown in Figs. (7) $-$ (8). In contrast, we have assumed the various values of $\triangle$ showing qualitative impacts
 on $\Lambda$BHDE. We conclude that the diagnostic tools are efficiently applicable to distinguished BHDE models. Furthermore, $S_{4}^{(1)}$ 
 can generally predict larger variation amongst the cosmic evolution of the BHDE in comparison to $S_{3}^{(1)}$. In this way we get an easier 
 approach to distinguish different theoretical models.
 
 \item 
In addition, as compared with $S_{3}^{(1)}$, $S_{4}^{(1)}$ and $\epsilon(z)$, CND pairs have substantially distinct evolutionary routes.
They are more effective in diagnosing different theoretical models of DE. It implies that CND will provide the state with a great supplement
At the same time, the statefinder hierarchy and the cosmological data of structure growth rate were obtained by the CND process. 
 	
 \item 
 Thus the above study concludes that the BHDE and $\Lambda$CDM models can be easily differentiated by considering state finder for different 
 parametric values.
 	
 	.
 \end{itemize}  
  In summary, by using these diagnostic methods, we can conclude that the BHDE model can be easily differentiated. 
  
\section*{Acknowledgments}

The authors are thankful to Dr. Kasturi Sinha Ray (GLA University) for her help in preparing the manuscript.

\end{document}